# Wearing a single DNA molecule with an AFM tip

Sergio Santos[1,3], Victor Barcons[2], Josep Font[2], Neil H Thomson[1]

[1]School of Physics and Astronomy & Department of Oral Biology, Leeds Dental Institute, University of Leeds, LS2 9JT, UK. [2]Departament de Disseny i Programació de Sistemes Electrònics, UPC - Universitat Politècnica de Catalunya Av. Bases, 61, 08242 Manresa, Spain.

[3]Laboratory for Energy and NanoScience (LENS), Institute Center for Future Energy (iFES), Masdar Institute of Science and Technology, Abu Dhabi, UAE

**Abstract**

While the fundamental limit on the resolution achieved in an atomic force microscope (AFM) is clearly related to the tip radius, the fact that the tip can creep and/or wear during an experiment is often ignored. This is mainly due to the difficulty in characterizing the tip, and in particular a lack of reliable methods that can achieve this *in situ*. Here, we provide an *in situ* method to characterize the tip radius and monitor tip creep and/or wear and biomolecular sample wear in ambient dynamic AFM. This is achieved by monitoring the dynamics of the cantilever and the critical free amplitude to observe a switch from the attractive to the repulsive regime. The method is exemplified on the mechanically heterogeneous sample of single DNA molecules bound to mica mineral surfaces. Simultaneous monitoring of apparent height and width of single DNA molecules while detecting variations in the tip radius R as small as one nanometer are demonstrated. The yield stress can be readily exceeded for sharp tips (R<10 nm) at typical operating amplitudes (A>10nm).  The ability to know the AFM tip radius *in situ* and in real-time opens up the future for quantitative nanoscale materials properties determination at the highest possible spatial resolution.

**I. INTRODUCTION**

Initially, the AFM was developed to operate in contact mode (CM AFM) [1] whereby the tip is permanently contacting the sample during scanning and inducing relatively high lateral or frictional forces. With the introduction of dynamic modes, and, in particular, Tapping Mode (TM) or intermittent contact amplitude modulation (IC AM) AFM, samples are subjected to compressive forces while the frictional forces are highly reduced [2,3]. Nevertheless, even in tapping mode, tip-sample forces can still be too large, especially if high resolution is required [3-6]. High resolution invariably involves a small tip radius [6] which, in turn, results in low adhesion forces [7,8]. Ultra-sharp tips, however, may involve such high pressures in the contact region that tip wear or damage cannot be prevented [9]. Thus a compromise is needed to prevent damage to the tip and also the sample, especially when imaging soft matter. In this respect, San Paulo and Garcia [10] reported irreversible damage in the structure of IgG antibodies when the AFM was operated in the repulsive force regime (through the High (H) state) as compared to the attractive force regime (through the Low (L) state). Note that these make reference to the two stable states of oscillation that can co-exist when a cantilever oscillates near a surface [11-13] and that they are characterized by producing the same tapping amplitude at two different cantilever-sample separations; the L state typically involving the cantilever vibrating several nm higher above the sample than the H state [14]. It is unclear whether the changes reported by San Paulo and Garcia resulted from plastic deformation of the molecules or the tip or both. Round and Miles [15] conducted similar investigations on double-stranded DNA (dsDNA) and reported no plastic deformation occurring to the DNA in the L nor in the H state. It was suggested that differences in sample stiffness between protein and DNA might have caused the antibodies to be damaged in the H state while the DNA was able to withstand the forces induced by the intermittent contact. Moreover, Thomson [16] later resolved IgG antibodies in the repulsive force regime through the H state with no apparent molecular damage. Here we propose that these apparently divergent outcomes can be the result of differences in tip sharpness, and consequently, different pressures between the tip and the sample during intermittent contact and tip wear[9].

The capabilities of the AFM to study the mechanisms of friction and wear in the nanoscale have been realized for a long time [17]. Several groups have demonstrated tip wear and estimated the tip radius of an AFM probe via AFM and/or Scanning Electron Microscopy (SEM) images of the tip [18] and/or by measuring topographical features and fitting the data into theoretical models [19,20,21]. These methods all involve tip-sample contact during the characterization and typically involve scanning and scratching samples[22] or simply scanning micron-sized or high aspect ratio nanoscale features[23]. However, these methods can be destructive to the tip if the tracking force is not rigorously controlled. Other disadvantages of the above methods are that these might be time consuming, might require changing the sample or removing the tip from the AFM holder and, in some cases, might not be suitable to detect very small variations in tip radii when the tip is very sharp (e.g. R< 5-10nm)[24]. Thus, we propose a straight forward *in situ* method of determining the tip size. This method allows us to monitor creeping of the tip *in situ*, and in the more extreme cases, wearing of the tip and the wear of biological molecules such as DNA in ambient AM AFM. To this end, the inherent bi-stability of the oscillating cantilever near the surface is used to indirectly monitor the state of the tip. In short, the method has the potential to detect very small changes in tip radius (ΔR≤1nm) by monitoring changes in the behavior of the cantilever dynamics[24].

## II. RESULTS AND DISCUSSION

### A. Model

The dynamics of a cantilever in AM AFM with a high quality factor (Q) can be approximated well by the following equation of motion for a driven harmonic oscillator,

$$m\frac{d^2z}{dt^2} + \frac{m\omega_0}{Q}\frac{dz}{dt} + kz = F_{ts} + F_0 \cos \omega t \qquad (1)$$

where details regarding the definitions of variables and the validity of using a point mass model in ambient AM AFM can be found in the literature [25-27]. In the present work (1) has been solved numerically with the use of commercially available software[28]. The Dejarguin-Muller-Toporov[8] (DMT) model of contact mechanics has been used for the contact and adhesion force and the van deer Waals force for the long range as detailed in Ref. 26. The parameters for the simulations are k=40N/m (spring constant), $f_0$=300 kHz (natural resonant frequency), $E_t$=70 GPa (elastic modulus of the tip), $E_s$=1.5 (elastic modulus of the sample), ν =0.3 (Poisson's coefficient), H =7.1x10$^{-20}$J (Hamaker constant), γ=35mJ/m$^2$ (surface energy) and Q=500 (Q factor). All experiments and simulations have been carried out at the free resonant frequency of the cantilever. Since the method depends on a basic knowledge of previously reported theoretical results, a brief summary is given here. Firstly, the L state typically involves non-contact imaging provided small enough free amplitudes are used[13,26]. Secondly, the average tip-sample interaction force might be net attractive or net repulsive. These two regimes are the so-called attractive and repulsive regimes and typically correspond to the L and H state when these exist. Thirdly, the phase shift provides an immediate experimental method of verifying whether the cantilever is oscillating in the L or the H state[26]. Thus, in what follows, the L and the H state will be differentiated according to the phase shift; values above (below) 90° correspond to the L (H) state respectively; phase values are given in all the experimental scans presented in this work. Note that, detecting whether the oscillation occurs in one or the other state however does not strictly depend on the phase convention but on phase behavior. This implies that even when the phase convention fails in terms of determining force regimes, such as at low set-point amplitudes, differences in phase between the L and H states are still large enough to allow distinguishing between them[29]. Fourthly, decreasing the amplitude set-point and/or increasing the free amplitude increases the probability of reaching the H state [14,30]. A situation in which the L state is inhibited at small values of amplitude set-point ($A_{sp}/A<<1$) has also been reported[31]. However, it has been found experimentally and by SEM characterization that this only occurs when relatively blunt tips (R>20-30nm) and/or compliant cantilevers are used (e.g. k=2N/m), at least for relatively stiff surfaces such as mica[29]. Since only intermediate to high set points ($A_{sp}/A \geq 0.30$) and very sharp tips are used experimentally in this work, the fourth point applies throughout.

Here, the pressure in the tip-sample interface has been calculated with the use of the continuum DMT theory of contact mechanics[32] which gives the contact area $a_{DMT}$ (2) as a function of indentation δ and tip radius R.

$$a^2_{(DMT)} = \delta R \qquad (2)$$

It is also useful[32] to introduce the concept of mean normal pressure $p_m$ (3) acting between the tip and the sample.

$$p_m = \frac{4}{3}\frac{E^*}{\pi}\sqrt{\frac{\delta}{R}}$$

(3)

$$\sigma_{N(MAX)} = -\frac{3}{2}p_m \qquad\qquad r=0 \qquad (4)$$

Note that the maximum normal pressure (or stress), i.e. the stress at the center of contact (r=0), can be expressed in terms of the mean normal pressure as shown in (4). The negative sign accounts for compression. From (4) the maximum value of stress is only 1.5 times the mean value. The results of simulating force and stress distance curves at resonance using (1) are shown in Fig. 1 as a function of tip radius and free amplitude; only the behavior while approaching the surface is shown for clarity. The plots show both average and peak values of force (top row) (both maximum or repulsive and minimum or attractive) and mean pressure (bottom row) per cycle. It is reasonable to expect that peak forces and pressure are more relevant in terms of wear and irreversible deformation than average values[9]. In fact, it can be argued that the scale shown in Fig. 1 is not appropriate to distinguish between average forces for the different values of R. Nevertheless a magnified plot is redundant since it is only for very small values of cantilever-sample separations that the average values of force show a slight dependency on tip radius (data not shown). Thus, average forces have been drawn with continuous lines for all values of A and R. It can be readily observed that that peak values can be up to ten times higher than average values. Additionally, as recently reported, pressure values, both peak and average, show a much stronger dependence on tip radius than forces[9]. Also note that for A=10nm there is a switch to the H state only when R=6 nm. This is a general characteristic of a tip vibrating over a surface where the free amplitude needs to be gradually increased to reach the H state as the tip radius increases. This fact has long been known and reported by several groups[25,33]. Nevertheless, this is generally ignored when interpreting experimental data probably due to the difficulties of estimating the tip radius *in situ* and due to the other dependencies that such an effect might have. Significantly, for the smallest values in Fig. 1 (A=10nm, R=6nm), the average and peak forces in the H state are much lower than those obtained for A=20-30nm for all values of R (Fig. 1a-c). Nevertheless, the peak and average pressure is as high or even higher for A=10nm and R=6nm than it is for A=20-30nm when R≥10nm (compare Fig. 1d with f). The physical interpretation is that neither free amplitudes nor average or peak forces provide sufficient criteria to establish whether the interaction is soft[9].

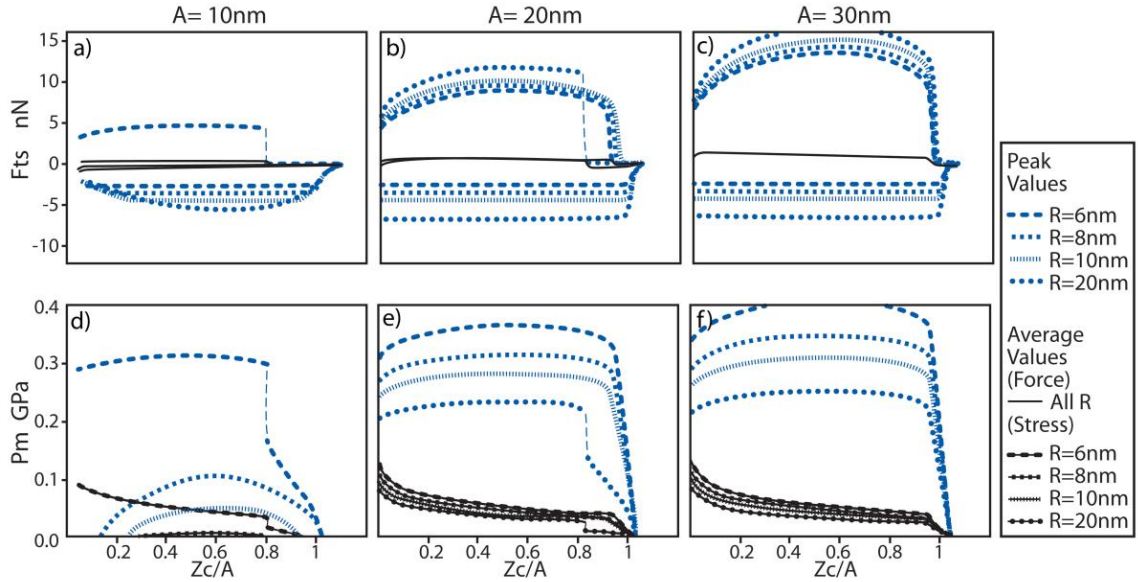

FIG 1. Relationships between average and peak forces, and average and peak stress per cycle as a function of normalized tip-sample separations $z_c/A$ for a range of free amplitudes and tip radii. In the top row, average and peak forces are shown for A = (a) 10nm, (b) 20nm and (c) 30nm. Both the maximum (repulsive) and minimum (attractive) peak forces are shown. The average force values are also shown, represented by continuous lines for all tip radii for simplicity since these almost completely overlap. The respective average and peak stresses per cycle are shown in the bottom row in (d), (e) and (f). In the case of A=10nm, the H state is only reached when the tip radius is as small as R=6nm, nevertheless some contact also occurs in the L state for R=6, 8 and R=10nm as deduced from the peak stresses in (d). Average forces range from approximately 0.2 nN in (a) to 1.5 nN in (c).

In summary, from the above data and also from previous studies where (1) has been solved numerically [14,25,26,33], four relevant concepts can be deduced in order to monitor the broadening of the tip and molecular wear *in situ*. Firstly, even relatively small values of free amplitude, and therefore, average and peak forces, can induce high or very high pressure in the H state if the tip radius is sufficiently small[9]. However, stress rapidly falls with increasing tip radius even when using relatively much higher free amplitudes (Fig. 1)[9]. Note that the yield stress of even high-strength materials ranges from 0.3 to 1GPa [34], so the model predicts that tip-sample pressures come into the range expected of plastic deformation, particularly for sharp tips (R<10nm). Secondly, it is not reasonable to expect the pressure to indefinitely increase. Therefore, for a given tip radius, it is expected that there will be a critical value of free amplitude for which the tip will either start to creep or wear as a mechanism to reduce pressure in the interaction[9]. Thirdly, as the tip wears or creeps and as the tip radius increases, higher free amplitudes and/or smaller amplitude set-points will be required to reach the H state for a given cantilever-sample system[14,24,30]. Finally, while surface contacts are, in general, not smooth but complex, and consist of asperity distributions that can deform both elastically and plastically when submitted to contact [17,32,35], for simplicity, in the model, both the sample and the tip have been assumed to be smooth and deform only elastically as the load is applied. While this last assumption is not realistic if very high pressures are induced, the model still gives realistic indications of the relationships between operational parameters and the pressure in the tip-sample interface. Moreover, the predictions from the model are validated in our experiments[24] as it will be discussed next.

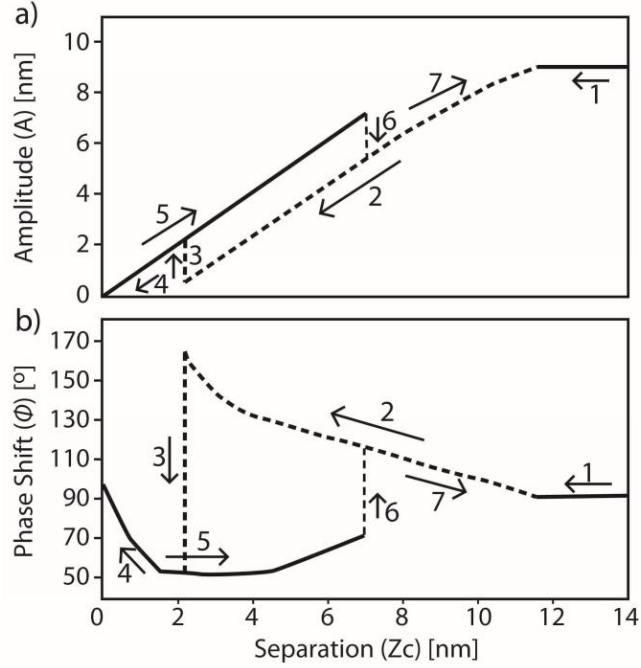

FIG 2. (a) Amplitude-distance curve performed on a mica surface at 40% RH for a silicon rectangular cantilever with nominal spring constant of 40 N/m driven at resonance ($f_0$=317kHz) with a free amplitude A=9nm. (b) Corresponding phase-distance curve where a step-like discontinuity is observed as the cantilever goes from the L to the H state when approaching the sample and from the H to the L state during retraction. The path followed is marked from 1 to 7, for the cantilever approach and retraction.

It is interesting to note that the mechanism for reducing pressure while imaging, i.e. tip broadening or changes in R[9], might affect the values of apparent height of nanostructures[36] and, in general, greatly affect the dynamics of the cantilever by inducing changes in the value of $A_c$[24,37]. This can be readily observed from Fig. 1 while previous studies[33,38] have also indicated a direct dependency of $A_c$ on R. In fact, it can be shown experimentally[24] that a power law can be found for this relationship. In particularly, for an Olympus AC160TS cantilever and a mica surface, one can obtain the relationship $R = 4.75A_c^{1.12}$. Typical power law expressions that can be derived this way can have errors for R of 1-2 nm or less[24]. The above law will be used next to show how the creep and/or wear of a tip and the wear of a dsDNA molecule can be monitored *in situ*. Monitoring the tip radius *in situ* can be used to, for example, accurately predict the loss in apparent height of nanostructures due to the finite size of the tip[36], retrieve quantitative information from the sample[37], establish whether the tip radius is stable during a given sequence of experiments and even provide quantitative information about the hydrophilicity of single molecules[39].

**B. Experimental demonstration of single molecule wear and tip creep *in situ***

For an Olympus AC160TS cantilever used on a mica sample, $R = 4.75A_c^{1.12}$ predicts that for $A_c \approx 6$ nm, R≈3±2 nm follows. Olympus AC160TS cantilevers have typical resonant frequencies of 300 kHz and nominal spring constants k≈40 N/m . The nominal value of tip radius is quoted to be R≈10±1 nm according to the manufacturer. Fig. 2 shows a typical APD curve obtained for the above cantilever model on a mica sample with A≈9nm at resonance. A switch to the H state is observed. This small value of $A_c$ can be readily interpreted as an indication of tip sharpness[24]. In particular, for this experimental set-up it was found that $A_c \approx$ 6 nm and, as stated, from $R = 4.75A_c^{1.12}$, it follows that R≈3±2 nm. This value of $A_c$ was obtained by performing a sequence of APD curves as described elsewhere[24,33] and shows how the combination of the experimental value of $A_c$ and a suitable power law, i.e. $R = 4.75A_c^{1.12}$, for a given cantilever-sample system can be used to rapidly characterize the tip radius *in situ*. Demonstration of the monitoring of tip creep and/or wear and single molecular wear of dsDNA *in situ* follows.

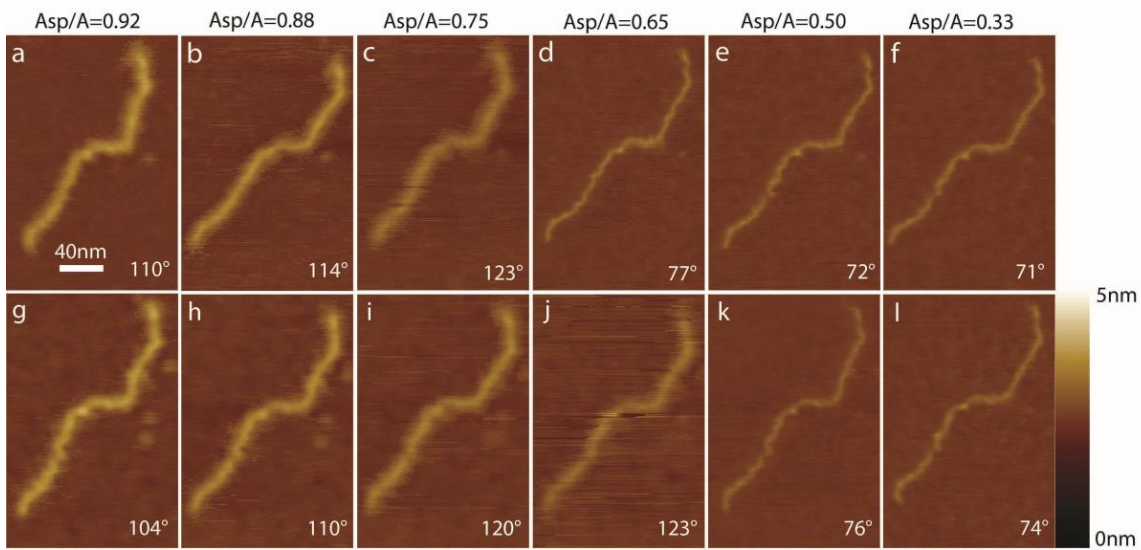

FIG 3. Two consecutive sequences, (a) to (f) and (g) to (l), of topographic images of a single 800bp dsDNA molecule on mica from which tip wear can be readily deduced. The average phase shift for every scan is shown on the bottom right in every image and allows differentiating between the L and the H state (A=9nm). R≈3 ±2 nm in a to f and R≈3.5 ±2 nm in g to l.

Fig. 3 shows a sequence of topographic images of a single DNA molecule on mica obtained with the same cantilever used to acquire the APD curve shown in Fig. 2 immediately after the acquisition of the curve. From a to f, the normalized amplitude set-point ($A_{sp}/A$) has been systematically reduced from 0.92 to 0.33 and the H state has been stably reached at intermediate values of set-point (Fig. 3d) as predicted by the APD curve in Fig. 2. Both apparent molecular height and width display dramatic differences in the L state compared to the H state. It is important to note however that heights of single molecules in AFM do not necessarily correlate with deformation[36]. Also note that the noise observed (Figs. 3b-c and 3g-j) is not a consequence of an inappropriate choice of feedback gains but a consequence of switching between states[14]. Gain optimization in terms of stability is an important issue in AFM in general [40], but in this situation no choice of gains makes the noise disappear. Significantly, a second sequence of scans (Fig. 3g-l) shows that the H state could not be reached stably for a second time with $A_{sp}/A=0.65$ (Fig. 3j) but had to be further reduced to 0.50 (Fig. 3k) in this case. This effect readily indicates that creep or wear of the tip has already occurred as a consequence of reaching the H state in Figs. 3d-f. At this point APD curves showed that $A_c \approx 7$ nm, and, from $R = 4.75 A_c^{1.12}$, it follows that R≈3.5 ±2 nm. After the scan in Fig. 3, over 20 scans of the same molecule were acquired with 2< A<9nm and for a whole range of amplitude set points (0.3<$A_{sp}$/A<0.92) and no further change in the cantilever dynamics were observed (data not shown). This readily indicates that the tip had stabilized for these relatively small values of A during the experiments. This behavior is general, has been routinely reproduced (data not shown) and it is consistent with our previous studies[9,13,37]. These results also indicate that it is possible to stably acquire tens of scans of the same molecule with relatively high resolution, and with the use of stiff cantilevers, without causing tip creep or wear once a critical broadening of the tip has been reached. This is only true, provided the free amplitude is kept below a certain critical value. Nevertheless, sharp tips, of say R<10 nm, tend to slowly creep or wear and eventually broaden according to our experiments even if the free amplitude is not increased above the current $A_c$ value for the system. It is also worth pointing out that we have only obtained high lateral resolution, comparable to Fig. 3d, whenever the H state has been reached with small values of free amplitude, i.e. $A=A_c<10$nm. This confirms the validity of the concept of $A_c$ to predict tip sharpness[24].

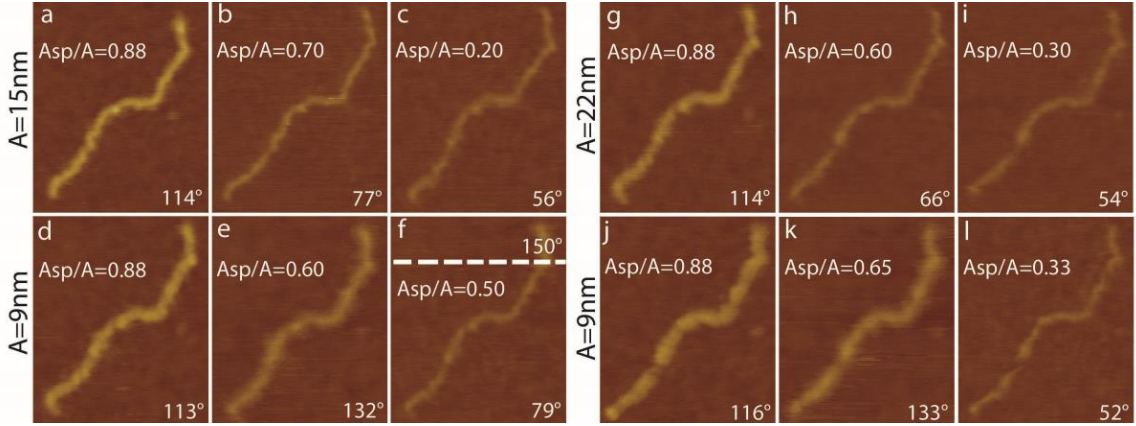

FIG 4. (a)-(c) Sequence of topographic images for which broadening of the tip and molecular wear are induced with A=15nm. (d)-(f) Control sequence with A=9nm where permanent damage to the molecule can be observed and slight tip creep can be deduced. In (f), a switch from the H state back to the L state, as indicated by the dashed line, is observed, showing that the H-state is no longer stable at $A_{sp}/A=0.50$. The same experiment shows that increasing A to (g)-(i) 22nm results in (j)-(l) further tip broadening and molecular irreversible damage. R≈3.5 ±2 nm in a to c, R≈4 ±2 nm in d to i and R≈4.5 ±2 nm in j to l.

The tendency of the tip to broaden as A is slowly increased from 9 nm to 15 nm and then 22 nm is shown in Fig. 4. After scanning the molecule with A=15 nm (Fig. 4a-c) the three control scans obtained with A=9nm (Figs. 4d-f) show that the H state is no longer stable with $A_{sp}/A$=0.50 (Fig. 4f); note a switch back to the L state at the top of the scan. Here, APD curves showed that $A_c$≈8 nm, and, from $R = 4.75 A_c^{1.12}$, one obtains R≈4 ±2 nm. The increase of 0.5 nm in R relative to the state of the tip in the scans in Fig. 3g-l is directly attributable to the increase in A in the scans in Figs. 4a-c. Further increasing A to 22 nm (Figs. 4g-i) produces visible irreversible damage to the DNA molecule as observed in the control scans (A=9nm) in Figs. 4j-l. In this case the H state could not be reached with $A_{sp}/A$=0.50 (data not shown) any longer and it had to be further reduced to $A_{sp}/A$=0.33 (Fig. 4l) in order to stably image in the H state. This implies that the tip has broadened at the same time as molecular damage has occurred. Here, again, APD curves showed that $A_c$≈9 nm, and, from $R = 4.75 A_c^{1.12}$, R≈4.5 ±2 nm. The further increase in R of 0.5 nm is, again, directly attributable to the increase in A when performing the scans in Figs. 4g-i. In order to demonstrate that the effects observed in Figs. 4j-l are due to irreversible molecular damage, a nearby molecule on the same sample has been imaged with the control parameters (A=9nm) (Fig. 5). It is readily observed that while the dynamics of the cantilever are reproducible as compared to Figs. 4j-l, the contrast in both the L and the H states is sharper in Fig. 5 (see Fig 4. and compare these in Fig. 6). Furthermore, the molecule in Fig. 5 does not display discontinuities along the length of the molecule, symptomatic of localized damage as in Figs. 4j-l.

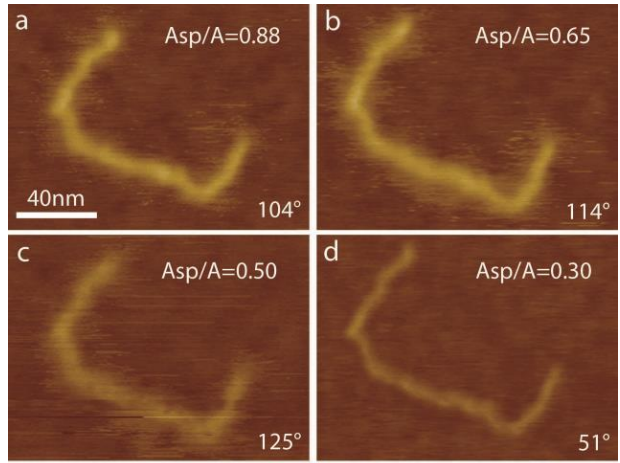

FIG 5. Sequence of topographic images of a nearby molecule reproducing a sequence similar to Fig. 3 where $A_{sp}/A$ is systematically reduced. The relatively high values of apparent height are almost completely recovered as compared to Fig. 3 in (a)-(b) the L state but the amplitude set-point has to be lowered to (d) $A_{sp}/A<0.30$ (A=9nm) in order to reach the H state stably. This is due to tip broadening relative to Fig. 3a-f where R≈3 ±2 nm. Here, R≈4.5 ±2 nm.

This sequence of experimental data taken with the same AFM tip shows how to monitor incremental broadening of the tip and how molecules can be slowly damaged when scanned with sharp tips and gradually increasing A. Nevertheless, our results show that dramatic and immediate irreversible damage of both tip and sample should be expected if a sharp tip is suddenly engaged with relatively high free amplitudes (A>20-30nm). For example, these cantilevers with R<10nm can completely destroy a DNA molecule if engaged with a free amplitude as low as 30-40nm and a set-point as high as $A_{sp}/A>0.80$ when trying to reproduce the above experiments (e.g. with scan sizes of 200-400nm)[9]. In fact, it can be verified experimentally that it only takes a single scan with these parameters to greatly broaden the tip to values as high as R>20nm ($A_c$>35nm)[9,24].

However, once high values of $A_c$ are reached (e.g. $A_c$>30-40nm), the H state can be safely reached with values of A typically used in tapping mode (10nm<A<60nm) without producing apparent molecular damage. This is in agreement with the relationships shown in Fig. 1 between pressure and tip radius and also agrees with other recent studies[9,13]. With regard to cantilever stiffness, we have also experimentally verified both that the tip radius can significantly broaden and that DNA molecules can be greatly damaged even when using values as low as k=2 N/m. This is true provided sharp tips, i.e. R<5-10nm[24], are used to reach the H state; note that higher values of $A_c$ are required to reach the H state with compliant cantilevers[13,24]. Finally, Fig. 6 shows the minimum, average and maximum apparent a) width and b) height of every scan shown in this article and allows for numerical comparison. Squares, circles and triangles stand for L state (attractive), H state (repulsive) and bi-stability (B) respectively. First note that in the example shown in this article, the gap between states is of approximately 2 nm as shown in the amplitude curve (Fig. 2). The importance of the tip-sample gap in terms of resolution and sensitivity has long been acknowledged [4]. More recently, the dramatic increase in resolution with even small increments in cantilever-sample distance (<1nm) has been demonstrated in Frequency Modulation (FM) AFM by atomically resolving an absorbed pentacene molecule[41]. Furthermore, it has been shown that the apparent height can be readily affected by the dynamics of the cantilever and, in particular, by the mode of operation used to scan the samples[36]. In this respect, the differences in height and width observed in Fig. 6 between the H and the L state, especially when the molecules are undamaged and the tips are sharp (compare 3a and 3d), experimentally demonstrate the relevance of the tip-sample gap in terms of apparent height generation in AM AFM. That is, these provide evidence of the dependency of apparent height and lateral resolution on force regimes, i.e. the average tip-sample distance, especially when the tips are sharpest, i.e. R≈3 ±2 nm in Figs. 3a-f. Furthermore, dramatic differences in apparent height of dsDNA molecules have been reported when using tips of R≈5 nm as compared to tips of R≈30 nm[36]. The minimum effects on apparent height with small variations in tip radius can be observed, for undamaged molecules, when comparing Figs. 3a-f to Figs. 5a-d (see Fig. 6) and nothing that the tip radius varies there approximately 1.5 nm only.

Several other patterns can be deduced from Fig.6 and by taking into account the value of R as predicted for these scans. First, the best lateral resolution (approximately 6nm in apparent width) occurs during the first scan sequence (Fig. 3d-f) in the H state when the tip is sharpest; $R = 4.75 A_c^{1.12}$ predicts R≈3-3.5 ±2 nm. For the second scan sequence shown in Fig. 4a-f, $R = 4.75 A_c^{1.12}$ predicts R≈3.5-4 ±2 nm and the apparent width is seen to slightly increase in this case (see Fig. 6). This is a general characteristic according to our results. That is, as $A_c$ increases, the tip radius is predicted to broaden according to $R = 4.75 A_c^{1.12}$ and the apparent width, i.e. resolution, of molecules in the H state (repulsive regime) worsens. Second, if biomolecules are imaged with sharp tips with relatively high values of free amplitude, i.e. $A_0>3/2A_c$, molecular damage readily occurs (see Figs. 4 and Figs. 6e-l). Third, provided the tip radius is relatively sharp and molecular damage has not occurred, the apparent height in the attractive regime (L state) should be highest. This agrees with our recent findings[36] and with the readings in Fig. 6. Fourth, provided the tip stays relatively constant during scans, apparent height readings of nanostructures, and, in particular, biomolecules, should be reproducible (albeit compromised by a finite tip size){Santos, 2011 #169}. This can be readily observed when comparing the apparent heights of the first biomolecule in Fig. 4 and in the L state when it is undamaged (Fig. 6a) with the second molecule (undamaged) in Fig. 5 (Figs. 5a-b). Recall that the tip radius has increased only by approximately 2 nm during the scans and thus, variations in apparent height due to tip broadening should be minimized. This is consistent with the above discussion.

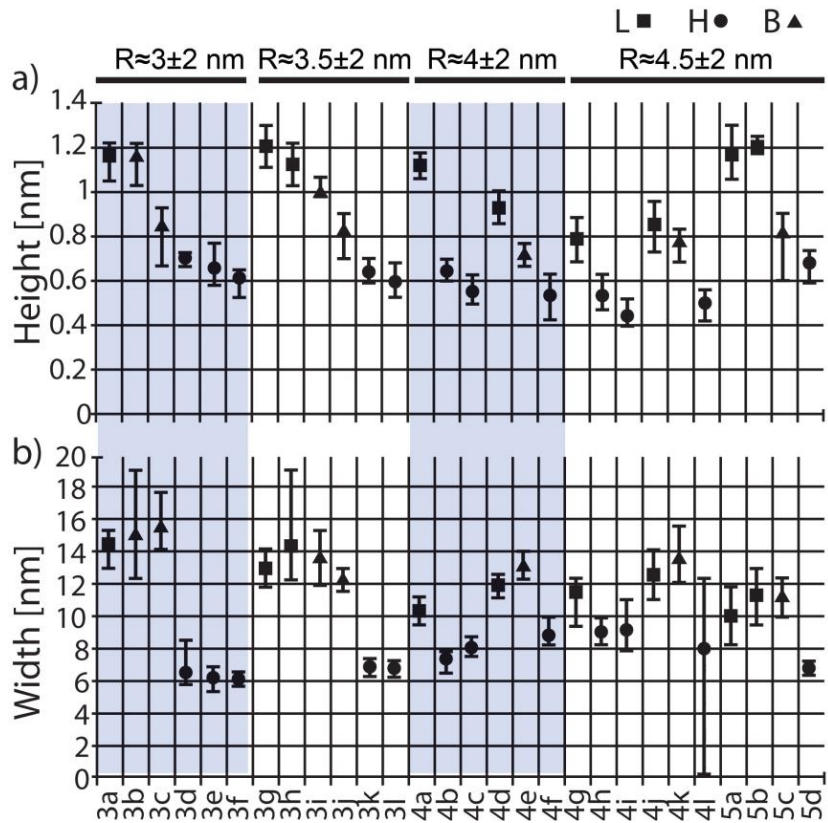

FIG 6. (a) Apparent height and (b) width of dsDNA for all the scans shown in this work taken with the same tip. The L and H states and bi-stable behavior are differentiated by squares, circles and triangles respectively. The tip radius estimated for each sequence of scans is written at the top of each sequence. The markers from 3a to 4l correspond to the first molecule probed with the tip in Figs. 3 and 4 and the markers from 5a to 5d

correspond to the second molecule probed in the scans in Fig. 5. Comparisons in terms of the two undamaged molecules can be made for 3a to 3f for the first molecule and from 5a to 5d for the second molecule. When the molecules are undamaged, changes in the dynamics of the cantilever due to slight creep and/or wear of the tip should account for differences in apparent height and width in the L (attractive) and H (repulsive) states. Since only slight variations of tip radius, R, have occurred during the scans, the differences in apparent height are minimal.

## III. CONCLUSION

A systematic approach to investigating the wearing of the tip of an AFM and a single molecule in AM AFM in ambient conditions has been provided. It has been shown that it is possible to distinguish between the wear of the tip and single molecules *in situ* by using the inherent bi-stability characteristics of a cantilever vibrating near a surface. This has also allowed good estimation of the sharpness of the tip. It has also been shown that force regimes and oscillation states do not account, on their own, for lateral and topographic resolution or tip and sample plastic deformation: stress in the tip-sample junction and tip radius has to be considered as well. This approach can help avoid divergent results and interpretations of molecular features and dimensions when imaging soft matter, particularly, isolated biomolecules on hard support surfaces such as mica. In particular, the fact that San Paulo and Garcia reported irreversible antibody damage even with the use of relatively small free amplitudes (A<10nm) when imaging in the H state while Thomson reported stable imaging even with the use of higher free amplitudes (A>30nm), can be readily interpreted as a difference in tip radius and pressure as predicted in this work.

ACKNOWLEDGEMENTS

We would like to thank Nick Herbert for helpful discussions on the interpretation of classical models and their limitations with regards to quantum effects, and Tony Fischer for his help with the interpretation of the pressure distribution in the contact area and the possible mechanisms of wear. We would also like to thank Bill Bonass and Daniel Billingsley for kindly providing us with dsDNA. Sergio Santos was funded through a Doctoral Training Grant of the BBSRC with additional funding from Asylum Research Corporation.